\documentclass[%
superscriptaddress,
nofootinbib,
amsmath,amssymb,
aps,
prd,
]{revtex4-2}

\usepackage{graphicx}
\usepackage{dcolumn}
\usepackage{bm}
\usepackage[mathlines]{lineno}
\usepackage{hyperref}

\hypersetup{colorlinks=true,citecolor=green,linkcolor=red,urlcolor=blue}

\begin{document}

\preprint{APS/123-QED}

\title{\texttt{QuickGWecc}: Fast Bayesian pipeline for searching eccentric binaries in pulsar timing array data}

\author{Lankeswar Dey}
\email{lankeswar.dey@nanograv.org}
\affiliation{Institute of Astrophysics, Foundation for Research \& Technology -- Hellas, GR-71110, Heraklion, Greece}

\author{Bence B\'ecsy}%
\affiliation{Institute for Gravitational Wave Astronomy and School of Physics and Astronomy, University of Birmingham, Edgbaston, Birmingham
B15 2TT, UK}%

\author{Abhimanyu Susobhanan}%
\affiliation{School of Physics, Indian Institute of Science Education and Research Thiruvananthapuram, Maruthamala PO, Thiruvananthapuram, Kerala 695551, India}%

\author{Maria Charisi}%
\affiliation{Institute of Astrophysics, Foundation for Research \& Technology -- Hellas, GR-71110, Heraklion, Greece}
\affiliation{Department of Physics and Astronomy, Washington State University, 1245 Webster Hall, Pullman, WA 99164, USA}

\date{\today}

\begin{abstract}
Recent pulsar timing array (PTA) results have provided evidence for the presence of a nanohertz gravitational wave (GW) background, most likely originating from a population of supermassive black hole binaries (SMBHBs). 
The next major milestone is the detection of continuous GWs (CGWs) from an individual SMBHB and the identification of its electromagnetic counterpart.
Typically, searches for CGWs in PTA data assume binaries in circular orbits.
However, theoretical studies indicate that binaries emitting GWs in the PTA frequency band may retain significant eccentricity.
In this paper, we present a fast and efficient Bayesian pipeline to search for CGWs from individual SMBHBs in relativistic eccentric orbits in PTA datasets. 
Our approach extends the \texttt{QuickCW} framework for circular binaries, in which model parameters are divided into \textit{projection} and \textit{shape} parameters.
Computational efficiency is achieved by performing many updates of the projection parameters for a fixed set of shape parameters, an operation that is orders of magnitude faster than updating the shape parameters.
We validate the pipeline through both detection and upper-limit analyses of simulated PTA datasets. 
This framework makes the search for eccentric SMBHBs computationally feasible, even in the next-generation PTA datasets with many more pulsars.
\end{abstract}

\maketitle


\section{Introduction}
\label{sec:intro}

Supermassive black hole binaries (SMBHBs) were first proposed to reside at the centers of some active galactic nuclei (AGNs) as a natural outcome of galaxy mergers in 1980 \citep{Begelman+1980}.
While galaxy mergers are expected to be common in hierarchical structure formation, the subsequent formation and long-term evolution of bound SMBHBs, particularly at sub-parsec separations, remains poorly understood \citep{DeRosa+2019}.
Despite more than four decades of observational and theoretical efforts, direct observational confirmation of close (sub-parsec separation) SMBHBs remains elusive.
In principle, two complementary messengers---electromagnetic (EM) radiation and gravitational waves (GWs)---provide independent avenues to detect and characterize these systems across different evolutionary stages.

An AGN hosting an SMBHB may exhibit periodic or quasi-periodic variability in its light curve due to the orbital motion of the binary (see \cite{DOrazio&Charisi2023} for a review). 
Such signatures arise naturally from a variety of physical mechanisms, including relativistic Doppler boosting, gravitational self-lensing, and periodic modulation of accretion rate \citep{DOrazio+2015, LehtoValtonen1996, DOrazio&Charisi2023}.
These expectations motivated systematic searches in photometric AGN surveys, which have revealed hundreds of SMBHB candidates \citep[e.g.,][]{Sillanpaa+1988, Sudou+2003, Graham+2015, Charisi+2016, Dey+2018}.
However, none of these candidates has been conclusively confirmed as alternative astrophysical explanations for the observed variability remain viable \citep[see, e.g.,][]{Robnik+2024}.

Complementary to EM signatures, SMBHBs at sub-parsec separations are expected to emit GWs in the nanohertz (nHz) frequency band, making them accessible to pulsar timing array (PTA) experiments \cite{Jenet+2004, Burke-Spolaor+2019, Taylor2021}. 
PTA experiments precisely monitor the times of arrival (TOAs) of radio pulses from an ensemble of millisecond pulsars distributed across the sky \cite{Sazhin1978, FosterBacker1990}. 
Passing GWs induce characteristic, spatially correlated deviations in the TOAs of each pulsar, providing a distinctive observational signature \citep{HellingsDowns1983, EstabrookWahlquist1975}. 
In 2023, multiple PTA collaborations have independently reported strong evidence for a stochastic nHz GW background (GWB) in their datasets \cite{NANOGrav_2023_GWB, EPTA+InPTA_2023_GWB, PPTA_2023_GWB, CPTA_2023_GWB, MPTA_2025_GWB, AgazieAntoniadis+2024}. 
The most plausible origin of this GWB is the incoherent superposition of GWs originating from a population of SMBHBs \citep{EPTA2024_astro, NG15_2023_astro}, although alternative sources---such as cosmological phase transitions, inflationary relics, and cosmic strings---remain viable \citep{Afzal+2023_NG15_NP}. 
If the GWB is indeed produced by a population of SMBHBs, the loudest individual systems are expected to be resolved over the background and be detectable as continuous GW (CGW) sources \cite{Gardiner+2025, Petrov+2025}. 
Detecting such CGWs is therefore critical both for confirming the astrophysical origin of the GWB and for establishing the existence of sub-parsec SMBHBs.
This further opens up the possibility of multi-messenger observation of SMBHBs if the host galaxy can be identified in the localization volume \citep{Kelley+2019_MM_PTA, Petrov+2024, Veronesi+2026}.

Several studies have searched for CGWs in different PTA datasets, with most analyses assuming binaries in circular orbits \cite[e.g.][]{Aggarwal+2019_NG11_CW, Arzoumanian+2023_NG12p5_CW, Falxa+2023_IPTADR2_CW, Agazie+2023_NG15_CW, EPTA_2024_CGW, Zha0+2025_PPTADR2_CW}.
This assumption is motivated by the expected circularization of binary orbits as they lose energy and angular momentum through GW emission, and by the relative computational simplicity of searches based on circular binary models.
However, theoretical and numerical studies of SMBHB evolution indicate that the interactions of the binary with nearby starts and its circumbinary disk may induce significant orbital eccentricity as it enters the PTA frequency band \citep{Sesana2010, Sesana2013, Franchini+2024}.
Searches based on circular waveform models have lower sensitivity to eccentric systems and may miss detectable signals \cite{Taylor+2016}. 
In \citet{Agazie+2024_3C66B}, we developed a pipeline for \textit{targeted} searches of CGWs from eccentric SMBHBs in PTA data that uses \texttt{enterprise} \citep{enterprise}, the standard PTA analysis package. 
Targeted searches aim to detect CGWs from known SMBHB candidates where EM observations of the host galaxy is used to inform parameters such as the sky location, distance, and sometimes the orbital frequency of the binary \citep{NG11_2020_3C66B_targeted, NG15_2026_114_targeted}. 
In contrast, in \textit{all-sky} searches these are treated as free parameters, significantly increasing the computational cost of the analysis \citep{Charisi+2024}. 
The targeted search pipeline was applied to search for CGWs from an eccentric SMBHB in the candidate source 3C~66B using the NANOGrav 12.5-year dataset \citep{Agazie+2024_3C66B}.
However, this method is computationally expensive, and it is not well-suited for all-sky searches for CGWs from eccentric SMBHBs, especially in the upcoming large PTA datasets.

\citet{Becsy+2022} introduced \texttt{QuickCW}, a method for fast Bayesian inference of CGWs from circular SMBHBs in PTA data. 
The speedup over conventional Bayesian search pipelines is achieved by separating the binary parameters into two classes: the \textit{shape} parameters that determine the intrinsic morphology of the GW signal emitted by the SMBHB, and the \textit{projection} parameters that describe how the signal is projected onto the line-of-sight of each pulsar \citep{Becsy+2022}.
This decomposition enables the computationally expensive waveform and noise-dependent quantities associated with the shape parameters to be reused across many likelihood evaluations. 
Consequently, updates to the projection parameters require only inexpensive recomputation of the projection coefficients, making likelihood evaluations for these updates orders of magnitude faster than those involving the shape parameters.
Rather than sampling all binary, pulsar-term, and red-noise parameters simultaneously, \texttt{QuickCW} adopts a block-updating strategy in which the projection parameters are updated many times while keeping the shape parameters fixed.
This approach substantially improves the overall sampling efficiency without compromising the correctness of the Bayesian inference.
In this work, we extend this framework for SMBHBs in relativistic eccentric orbits and develop \texttt{QuickGWecc}, a fast and efficient Bayesian pipeline to search for CGWs from relativistic eccentric SMBHBs in PTA data.

The paper is structured as follows. 
In Sec.~\ref{sec:pta_signal}, we describe the PTA response to CGWs from an eccentric SMBHB, outline the evolution of orbital parameters for binaries in relativistic eccentric orbits, and present methods for efficient likelihood evaluation. 
In Sec.~\ref{sec:sim_results}, we present simulated eccentric CGWs signals and results from applying \texttt{QuickGWecc} to recover these signals.
Finally, we summarize our work and discuss its possible future applications and extensions in Sec.~\ref{sec:summary}.

\section{PTA signal model and likelihood calculation}
\label{sec:pta_signal}

In PTA analyses, we begin with the measured TOAs and a best-fit timing model for an ensemble of pulsars.
The timing model is a mathematical model that describes various deterministic processes affecting the observed TOAs, including the rotation of the pulsar, its sky location and proper motion, its binary motion if applicable, the interstellar dispersion, as well as the instrumental effects \citep{Hobbs+2006}.
The difference between the observed TOAs and the TOAs predicted by the timing model are known as the timing residuals.
Non-zero timing residuals may arise due to a range of astrophysical and instrumental effects, including small deviations in the timing model parameters, the slow stochastic wandering of the pulsar's rotational period known as spin noise, stochastic variations in interstellar dispersion and the solar wind, instrumental noise, etc.
In addition, GWs passing between the pulsars and the Earth perturb the spacetime through which the pulsar's electromagnetic signal travels, inducing deviations in the TOAs resulting in timing residuals correlated across different pulsars.
Assuming the timing residuals $\delta t$ to be much smaller than the rotational period of the pulsar, we may write them as a sum of several contributions:
\begin{equation}
    \delta t = \Delta_{\text{TM}} + \Delta_{\text{WN}} + \Delta_{\text{IRN}} + \Delta_{\text{GWB}} + s \,,
    \label{eq:PTA_signal}
\end{equation}
where the term $\Delta_{\text{WN}}$ denotes uncorrelated white noise associated with the telescope receivers and the measurement process and $\Delta_{\text{IRN}}$ represents the intrinsic red noise of the pulsar. 
The red noise is characterized by a power-law spectrum with amplitude $A_{\text{RN}}$ and a spectral index $\gamma_{\text{RN}}$ (see \cite{NANOGrav_2023_NB} for more details).

The GW contribution to the timing residuals is separated into two components. 
The contributions from an isotropic stochastic background (the $\Delta_{\text{GWB}}$ term in Eq.~\ref{eq:PTA_signal}) is modeled as a common red-noise process with a power-law spectrum, parameterized by an amplitude $A_{\text{GWB}}$ and spectral index $\gamma_{\text{GWB}}$, with Hellings-Downs (HD) spatial correlation between different pulsars \citep{HellingsDowns1983}. 
The term $s$ in Eq.~(\ref{eq:PTA_signal}) represents the timing delay induced by CGWs from an individual SMBHB.
The marginalized timing model term $\Delta_{\text{TM}}$ allows the timing model parameters to be shifted from their best fit values as required by inclusion of these additional terms. 
Following standard practice in PTA analyses, we fix the white noise parameters to their best-fit values obtained from individual pulsar noise analyses \citep{Agazie+2023_NG15_DR, NANOGrav_2023_GWB}.

Variations in the ionized interstellar medium introduce delays in pulsar TOAs through changes in the dispersion measure (DM), defined as the integrated free electron density along the line-of-sight. 
These DM variations produce a chromatic timing delay that scales as $\nu^{-2}$, where $\nu$ is the observing radio frequency. 
In PTA analyses, DM variations are typically modeled either using a piecewise-constant function (the DMX model \citep{Agazie+2023_NG15_DR}), where the DM is allowed to vary independently in discrete time windows, or as a stochastic red-noise  Gaussian process (GP) with a power-law spectrum \citep{Larsen+2026_NG15_CNM}. 
In the latter approach, the DM-induced delays are treated analogously to intrinsic red noise but with the appropriate frequency dependence.
For this work, we assume that DM variations are modeled using the DMX approach, which is incorporated as part of the pulsar timing model and comes into Eq.~(\ref{eq:PTA_signal}) through the term $\Delta \text{TM}$.
In the following subsections, we describe the modeling of the CGW signal $s$, the orbital evolution of eccentric binaries, and our method for fast likelihood evaluation.

\subsection{The PTA signal induced by CGWs from an eccentric binary}

The timing delay induced by CGWs from an individual SMBHB in a pulsar is given by \cite{Taylor+2016, Susobhanan+2020}
\begin{equation}
    s(t) = \begin{bmatrix}
        F_{+} & F_{\times}
    \end{bmatrix}
    \begin{bmatrix}
        \cos{2\psi} & -\sin{2\psi}\\
        \sin{2\psi} & \cos{2\psi}
    \end{bmatrix}
    \begin{bmatrix}
        s_{+}(t-\tau_{p}) - s_{+}(t)\\
        s_{\times}(t-\tau_{p}) - s_{\times}(t)
    \end{bmatrix}\,,
    \label{eq:CGW_delay}
\end{equation}
where $F_{+,\times}$ are the antenna pattern response functions that encode the geometric projection of the GW onto the pulsar line-of-sight and depend on the sky positions of the GW source and the pulsar, and $\psi$ is the GW polarization angle \citep{Taylor+2016}.
The $+$ and $\times$ symbols denote the plus and cross polarizations of GWs, respectively.
The contributions to $s(t)$ that depend on the current time $t$ through $s_{+,\times}(t)$ correspond to the so-called \textit{Earth term}, as they represent the effect of the GWs passing the Earth at the time of observation.
In contrast, the contributions that depend on an earlier time $t - \tau_p$, are referred to as the \textit{pulsar term}, since they encode the effect of the GWs passing the pulsar at the time of pulse emission.
The geometric time delay $\tau_p$ between the Earth term and pulsar term contributions is given by
\begin{equation}
    \tau_p = d_p\,(1 - \cos{\mu})/c\,,
\end{equation}
where $d_p$ is the distance to the pulsar, $\mu$ is the angle between the lines of sight to the pulsar and the GW source, and $c$ is the speed of light in vacuum.
The Earth term is common to all pulsars, while the pulsar term depends on the individual pulsar location and distance, and probes the binary at an earlier evolutionary stage.
Since typical pulsar distances are of $\mathcal{O}(\text{kpc})$, the pulsar term provides a view of the binary evolution approximately $10^3$ years prior to the present day.

The analytic expressions of $s_{+,\times}(t)$ induced by GWs from an SMBHB in a relativistic eccentric orbit are given by \citep{Jenet+2004, Susobhanan2023}
\begin{subequations}
\label{eq:S+x_A}
\begin{eqnarray}
s_{+}(t) &=& \frac{H(t)}{n(t)} \left[(c_{\iota}^2 + 1)(-{\cal P}(t)\sin{2\omega(t)} + {\cal Q}(t) \cos{2\omega(t)}) + s_{\iota}^2 {\cal R}(t)\right]\,,\\
s_{\times}(t) &=& \frac{H(t)}{n(t)} 2c_{\iota} \left[{\cal P}(t)\cos{2\omega(t)} + {\cal Q}(t) \sin{2\omega(t)}\right]\,.
\end{eqnarray}
\end{subequations}
Here, $H = (GM\eta x)/(c^2 d_L)$, $G$ is the gravitational constant, $M = m_1 + m_2$ and $\eta = m_1m_2/M^2$ are the total mass and symmetric mass ratio of the binary, respectively, ($m_1, m_2$ are individual BH masses), and $d_L$ is the luminosity distance to the source.
The parameter $x = \left(G M (1+k) n/c^3\right)^{2/3}$ is the dimensionless post-Newtonian expansion parameter, $n = 2\pi/P$ is the mean motion, $P$ is the orbital period of the binary, and $k$ denotes the rate of advance of pericenter of the binary \citep{DGI_2004}.
Further, $c_{\iota} = \cos{\iota}$ and $s_{\iota} = \sin{\iota}$, where $\iota$ is the inclination angle of the orbital plane with respect to the sky plane (i.e., the angle between the orbital angular momentum and the line-of-sight), and $\omega$ is the argument of pericenter.
The functions ${\cal P, Q,}$ and ${\cal R}$ are given by
\begin{subequations}
    \begin{eqnarray}
        {\cal P} &=& \frac{\sqrt{1-e_t^2} \,(\cos{2u} - e_t\cos{u})}{1 - e_t\cos{u}}\,,\\
        {\cal Q} &=& \frac{((e_t^2-2)\cos{u} + e_t) \sin{u}}{1 - e_t\cos{u}}\,,\\
        {\cal R} &=& e_t\sin{u}\,, 
    \end{eqnarray}
    \label{eq:PQR}
\end{subequations}
where $e_t$ is time eccentricity and $u$ is eccentric anomaly \citep{DGI_2004, Susobhanan+2020}.

As the binary emits GWs, it loses orbital energy and angular momentum, leading to a gradual inspiral. 
Consequently, the orbital frequency (and hence $n$) increases with time, while the eccentricity $e_t$ decreases. 
In addition, both the eccentric anomaly $u$ and the argument of pericenter $\omega$ evolve due to the orbital motion and relativistic pericenter precession. 
In the following subsection, we describe how we model the evolution of these time-dependent parameters, which are required to compute the CGW-induced timing delay in the TOAs.

\subsection{Evolution of orbital parameters of relativistic eccentric binary}

The orbital evolution of a relativistic binary in an eccentric orbit can be separated into two parts: conservative and reactive evolution. 
The conservative part, as the name suggests, does not change the orbital energy or angular momentum and primarily manifests as the usual Keplerian orbital motion modified by the relativistic advance of pericenter. 
In contrast, the reactive part arises from the loss of orbital energy and angular momentum due to GW emission, leading to a secular evolution of the orbital period and eccentricity.
To model the orbital evolution, we employ the post-Newtonian (PN) quasi-Keplerian formalism, which generalizes the Keplerian motion to incorporate relativistic corrections.
This approximation is sufficiently accurate for the present work, as PTA experiments are expected to detect SMBHBs during the early inspiral phase, long before the binary approaches merger.
In this approach, the evolution is governed by a set of four coupled differential equations \citep{DGI_2004, Susobhanan+2020},
\begin{subequations}
    \begin{eqnarray}
        \frac{dn}{dt} &=& \frac{1}{5} \left(\frac{GMn}{c^3}\right)^{5/3} n^2\,\eta\, \frac{(96 + 292 e_t^2 + 37 e_t^4}{(1 - e_t^2)^{7/2}}\,,\\
        \frac{de_t}{dt} &=& -\frac{1}{15} \left(\frac{GMn}{c^3}\right)^{5/3} n\,\eta\,e_t\, \frac{(304 + 121 e_t^2}{(1 - e_t^2)^{5/2}}\,,\\
        \frac{d\gamma}{dt} &=& k\,n\,,\\
        \frac{dl}{dt} &=& n\,,
    \end{eqnarray}
    \label{eq:dndtdedt}
\end{subequations}
where $l$ and $\gamma$ denote the mean anomaly and the periastron angle, respectively. 
These equations are accurate to leading quadrupolar (2.5PN) order.

Given the initial values of $n$, $e_t$, $l$, and $\gamma$ at a reference time $t_0$, the coupled differential equations (Eqs.~\ref{eq:dndtdedt}) can be solved to obtain the temporal evolution of the orbital parameters. 
In this work, we use the analytic solution presented in \citet{Susobhanan+2020} to compute this evolution efficiently.
Once the temporal evolution of $l$ and $e_t$ are known, the evolution of the eccentric anomaly $u$ is obtained by solving the PN Kepler equation,
\begin{equation}
    l = u - e_t\sin{u} + \mathfrak{F}_t(u)\,,
\end{equation}
where $\mathfrak{F}_t(u)$ is a periodic function of $u$ that first appears at the 2PN order.
The true anomaly $v$ is then given by
\begin{equation}
    v = 2\,\arctan{\left[\sqrt{\frac{1+e_\phi}{1-e_\phi}} \tan{\frac{u}{2}}\right]}\,,
\end{equation}
where $e_\phi$ is the angular eccentricity. 
The evolution of the argument of pericenter $\omega(t)$ is computed as
\begin{eqnarray}
    \omega &=& \phi - v \nonumber\\
           &=& \gamma + k(v-l) + \mathfrak{F}_\phi(u) \,,
\end{eqnarray}
where $\phi$ is the orbital phase and $\mathfrak{F}_\phi(u)$ is another periodic function of $u$ first appearing at the 2PN order.

We thus obtain all the time-dependent orbital parameters required in Eqs.~(\ref{eq:S+x_A}) and (\ref{eq:PQR}) to compute the CGW-induced timing delay $s(t)$ in the pulsar TOAs. 
For further details on solving the coupled differential equations and implementing the orbital evolution, we refer the reader to \citet{Susobhanan+2020, Agazie+2024_3C66B}.

\subsection{Fast likelihood calculation}

The temporal evolution of $\omega(t)$ does not depend on its initial value $\omega_0$, and this allows us to rewrite Eqs.~(\ref{eq:S+x_A}) by expressing $\omega$ as $\omega = \omega(t) + \omega_0$, where $\omega(t)$ denotes the evolution assuming $\omega_0 = 0$. 
With this decomposition, the CGW-induced timing delay for the $i$th pulsar, $s_i(t)$, can be written using Eqs.~(\ref{eq:CGW_delay}) and (\ref{eq:S+x_A}) as
\begin{equation}
    s_{i}(t) = \sum_{j=1}^{5} \sigma_{5(i-1)+j}(\theta, \phi, H_0, \iota, \omega_0, \omega_{p_i}, \psi) ~~ S^{5(i-1)+j} (t, \theta, \phi, M, \eta, e_0, f_0, l_0, 
        l_{p_i}, d_{p_i})\,,
        \label{eq:quickGWecc_delay}
\end{equation}
the coefficients $\sigma$ and filter functions $S$ are defined below.
The parameters $(\theta,\phi) = (\pi/2 - \text{decl.}, \text{R.A.})$ are the polar and azimuthal angle of the GW source denoting its sky location, $H_0 = (G M \eta x_0)/(c^2 d_L)$, and $d_{p_i}$ is the distance to the $i$th pulsar. 
The quantities $x_0$, $e_0$, $l_0$, and $\omega_0$ denote the values of the respective parameters at the reference time $t_0$, $f_0 = 1/P_0$ is orbital frequency at $t_0$, while $l_{p_i}$ and $\omega_{p_i}$ denote the mean anomaly and argument of pericenter for the pulsar term of the $i$th pulsar, evaluated at $t_0 - \tau_{p_i}$.
In principle, $l_{p_i}$ and $\omega_{p_i}$ can be determined from the other binary parameters if the pulsar distances are known precisely. 
However, current uncertainties in pulsar distance measurements are typically orders of magnitude larger than the GW wavelength ($\sim$pc), and that makes it impossible to phase-connect the Earth and the pulsar terms. 
Therefore, we treat these quantities as free parameters.

Following \citet{Becsy+2022}, we separate the model parameters into two groups.
Parameters that enter only through the coefficients $\sigma$—namely $H_0, \iota, \omega_0, \omega_{p_i}$, and $\psi$—are the projection parameters. 
The remaining parameters, which determine the functions $S$, are the shape parameters. 
This separation is key to enabling efficient likelihood evaluations.
The explicit expressions for $S$ and $\sigma$ are given by
\begin{subequations}
    \begin{eqnarray}
        S^{5(i-1)+1} &=& \frac{x(t)}{x_0\, n(t)} \Big[{\cal P}(t)\sin{(2 \omega(t))} - {\cal Q}(t)\cos{(2 \omega(t))} \Big]\,,\\
        S^{5(i-1)+2} &=& \frac{x(t)}{x_0\, n(t)} \Big[{\cal P}(t)\cos{(2 \omega(t))} + {\cal Q}(t)\sin{(2 \omega(t))}\Big]\,,\\
        S^{5(i-1)+3} &=& \frac{x(t_{p})}{x_0\, n(t_{p})} \Big[{{\cal P}(t_{p})\sin{(2 \omega(t_{p}))} - \cal Q}(t_{p})\cos{(2 \omega(t_{p}))}\Big]\,,\\
        S^{5(i-1)+4} &=& \frac{x(t_{p})}{x_0\, n(t_{p})} \Big[{\cal P}(t_{p})\cos{(2 \omega(t_{p}))} + {\cal Q}(t_{p})\sin{(2 \omega(t_{p}))}\Big]\,,\\
        S^{5(i-1)+5} &=& \frac{1}{x_0} \left[\frac{x(t_{p}) {\cal R}(t_{p})}{n(t_{p})} - \frac{x(t) {\cal R}(t)}{n(t)}\right] \,,
    \end{eqnarray}
\end{subequations}
where $t_{p} = t - \tau_{p_i}$, denote the pulsar-term time for the $i$th pulsar, and
\begin{eqnarray}
    \sigma_{5(i-1) + 1} &=& H_0 \Big[ (c_{\iota}^2 + 1) \cos{2\omega_0}(F_+^i \cos{2\psi} + F_{\times}^i \sin{2\psi}) - 2 c_{\iota} \sin{2\omega_0} (F_+^i \sin{2\psi} - F_{\times}^i \cos{2\psi})  \Big] \,, \nonumber\\
    \sigma_{5(i-1) + 2} &=& H_0 \Big[ (c_{\iota}^2 + 1) \sin{2\omega_0}(F_+^i \cos{2\psi} + F_{\times}^i \sin{2\psi}) + 2 c_{\iota} \cos{2\omega_0} (F_+^i \sin{2\psi} - F_{\times}^i \cos{2\psi})  \Big]\,, \nonumber\\
    \sigma_{5(i-1) + 3} &=& -H_0 \Big[ (c_{\iota}^2 + 1) \cos{2\omega_i}(F_+^i \cos{2\psi} + F_{\times}^i \sin{2\psi}) - 2 c_{\iota} \sin{2\omega_i} (F_+^i \sin{2\psi} - F_{\times}^i \cos{2\psi})  \Big] \,, \nonumber\\
    \sigma_{5(i-1) + 4} &=& -H_0 \Big[ (c_{\iota}^2 + 1) \sin{2\omega_i}(F_+^i \cos{2\psi} + F_{\times}^i \sin{2\psi}) + 2 c_{\iota} \cos{2\omega_i} (F_+^i \sin{2\psi} - F_{\times}^i \cos{2\psi})  \Big]\,, \nonumber\\
    \sigma_{5(i-1) + 5} &=& H_0\, s_{\iota}^2 (F_+^i \cos{2\psi} + F_{\times}^i \sin{2\psi})\,.
\end{eqnarray}

The log-likelihood for the full PTA signal model (Eq.~\ref{eq:PTA_signal}) is given by \cite{Taylor2021, Becsy+2022}
\begin{equation}
    \log{\mathcal{L}} = -\frac{1}{2} (\delta t-s|\delta t-s) - \frac{1}{2}\log \det(2\pi C)\,,
    \label{eq:logLikelihood}
\end{equation}
where $(a|b) = a^TC^{-1}b$, and $C = N + TBT^T$ is the TOA covariance matrix. 
Here, $N$ is the white noise covariance matrix, $T$ is the design matrix describing the timing model and red noise, and $B$ is the prior covariance of the corresponding parameters \cite{Taylor2021}.
Following \citet{Becsy+2022}, the log-likelihood can be rewritten as
\begin{equation}
    \log{\mathcal{L}} = -\frac{1}{2} (\delta t|\delta t) - \frac{1}{2}\log \det(2\pi C) + \sum_{k=1}^{5N_p} \sigma_k N^k - \frac{1}{2} \sum_{k=1}^{5N_p} \sum_{l=1}^{5N_p} \sigma_k \sigma_l M^{kl}\,,
    \label{eq:fast_logL}
\end{equation}
where $N^k = (\delta t| S^k)$ and $M^{kl} = (S^k|S^l)$, and $N_p$ is the number of pulsars in the array.
Evaluating these inner products $N^k$ and $M^{kl}$ is the most computationally expensive step in the likelihood calculation. 
Whenever the shape parameters, red noise parameters, or GWB parameters are updated, the quantities $N^k$ and $M^{kl}$ must be recomputed. 
However, for fixed values of these parameters, the likelihood depends on the projection parameters only through the coefficients $\sigma_k$, which are inexpensive to evaluate. 
Expressing the likelihood like this allows us to perform a large number of updates of the projection parameters for each update of the shape parameters, significantly accelerating the overall sampling by reducing the number of expensive likelihood evaluations.
A summary of the shape and projection parameters, along with their scaling with the number of pulsars $N_p$, is given in Table~\ref{tab:shape_proj_params}.

\begin{table}[h]
    \centering
    \caption{Shape and projection parameters and their scaling with the number of pulsars ($N_p$). The parameters $A^{\rm RN}_{p_i}$ and $\gamma^{\rm RN}_{p_i}$ represent the red noise amplitude and spectral index of the $i$th pulsar.}
    \begin{tabular}{c|c}
    \hline
    \hline
        Shape parameters & ~~~Projection parameters \\
        $(9 + 4\times N_p)$ & ~~$(4 + N_p)$\\
        \hline
        $\theta, \phi, M, \eta, e_0, f_0, l_0, 
        l_{p_i}, d_{p_i}, \gamma^{\rm RN}_{p_i}, A^{\rm RN}_{p_i}, \gamma_{\rm GWB}, A_{\rm GWB}$~~ & ~~$\iota, H_0, \psi, \omega_0, \omega_{p_i}$\\
        \hline
        \hline
    \end{tabular}
    \label{tab:shape_proj_params}
\end{table}

With the likelihood formalism in place, the parameter estimation is carried out using Markov Chain Monte Carlo (MCMC) methods to sample the posterior distribution of the model parameters.
Following \citet{Becsy+2022}, we employ a Metropolis-within-Gibbs sampler together with a block-update strategy that exploits the separation between shape and projection parameters.
In this approach, the projection parameters are updated a large number of times (usually $\mathcal{O}(10^3)$) before attempting an update of the shape parameters. 
Since projection parameter updates require only recomputation of the coefficients $\sigma_k$, these updates are computationally inexpensive and can be performed rapidly.

To improve the sampling efficiency of the computationally expensive shape parameter updates, we use a multiple-try MCMC scheme \citep{Martino_2018_MTMCMC}. 
When an update to the shape parameters is proposed, the proposed shape parameters are combined with $N$ (typically $N=10^4$) different sets of proposed projection parameters.
Among these $N$ combined set of parameters, one is selected based on some importance weights, and the proposed update is accepted or rejected using an acceptance probability that depends on the likelihood values at all $N$ trial points. 
See Sec.~III of \citet{Becsy+2022} for more details about how the importance weight and acceptance probability are calculated.
This ensures that the updates to the shape parameters are not rejected due to lack of an appropriate set of projection parameters with very little additional cost, as likelihood evaluation at different projection parameters is comparatively cheap.
This approach significantly improves mixing and sampling efficiency in the high-dimensional parameter space.

For proposing updates to the shape parameters, we use a combination of Fisher-matrix proposals, differential-evolution proposals, and prior draws \citep[see][and references therein]{Becsy+2022}.
We also divide the shape parameters into the following groups: the seven common binary parameters $(\theta,\phi,M,\eta,e_0,f_0,$ and $l_0)$, pulsar distances and pulsar-term mean anomalies $(d_{p_i}$ and $l_{p_i})$, and red-noise parameters ($A_{p_i}^{\rm RN}$s, $\gamma_{p_i}^{\rm RN}$, $A_{GWB}$, and $\gamma_{GWB}$).
At each step, we only update a group of the shape parameters or all of them.
For projection parameter updates, each parameter is proposed independently using an optimal jump scale determined from the local curvature of the likelihood, estimated through the second derivative along that parameter direction \citep[see][for more details]{Becsy+2022}.

Combining the PTA signal model for relativistic eccentric binaries, the efficient calculation of the orbital evolution, and the fast likelihood formalism described above, we develop the \texttt{QuickGWecc}\footnote{\url{https://github.com/lanky441/QuickGWecc}} pipeline. 
The pipeline provides a computationally efficient Bayesian framework for performing both all-sky and targeted searches for continuous GWs from eccentric SMBHBs in PTA datasets.
In the following section, we demonstrate the performance of \texttt{QuickGWecc} using simulated PTA datasets for both detection and upper-limit analyses.

\section{Simulations and Results}
\label{sec:sim_results}

In this section, we describe the procedure used to simulate PTA datasets containing white noise, intrinsic pulsar red noise, a stochastic GWB, and CGWs from a relativistic eccentric SMBHB. 
We then present the results obtained by applying the \texttt{QuickGWecc} pipeline to these simulated datasets for both all-sky and targeted searches.

\subsection{Simulating dataset}

To simulate PTA datasets containing a CGW signal from an eccentric SMBHB, we begin with the timing models (\textit{par} files) of 25 longest timed pulsars from the NANOGrav 15-year dataset \citep{Agazie+2023_NG15_DR}. 
To simplify the dataset and reduce the computational cost of the analyses, we remove the `DMX' and `JUMP' parameters from the timing models and generate idealized TOAs for a single receiver and single observing frequency with bi-weekly cadence and 15 year data span for each pulsar. 
We then add white noise with EFAC $=1$ and $\log_{10}\text{EQUAD} = -6.5$, and injected intrinsic red noise using the best-fit red noise parameters obtained from the real NANOGrav observations for each pulsar.
In addition, we include a stochastic GW background with $\log_{10} A_{\rm GWB} = -14.62$ at GW frequency of $\text{yr}^{-1}$, the value reported in \citet{NANOGrav_2023_GWB}, and a spectral index $\gamma_{\rm GWB} = 4.33$, expected value for an isotropic GWB coming from a population of circular SMBHB. 
Finally, we inject the PTA signal from a relativistic eccentric SMBHB into the TOAs and refit the timing models to obtain updated best-fit par files for all pulsars.

\subsection{All-sky search}

We first apply the \texttt{QuickGWecc} pipeline to perform all-sky searches on the simulated datasets, assuming no prior information about the binary. 
We consider two separate analyses: (i) detection analysis where a strong, clearly detectable eccentric CGW signal is injected into the data, and (ii) upper limit analysis with a CGW signal just below the PTA sensitivity threshold.
At present, \texttt{QuickCW} and its extension \texttt{QuickGWecc} do not support searches in the presence of a HD correlated GWB \cite{HellingsDowns1983}. 
Therefore, in all analyses presented here, we model the GWB as a common uncorrelated red-noise (CURN) process, neglecting the spatial correlations between pulsars. 
Nevertheless, a likelihood reweighting scheme can be used to approximately recover the posterior distributions corresponding to an HD-correlated GWB model \citep{Hourihane+2023, Agazie+2023_NG15_CW}.

\subsubsection{Detection analysis}

\begin{table}[h]
    \centering
    \caption{Injected values of the SMBHB parameters for the detection and upper limit analysis. The last column shows the prior used for the binary parameters for both the detection and upper limit analysis.}
    \begin{tabular}{c|c|c|c}
    \hline
    \hline
        Parameter & Detection analysis & Upper limit analysis & Prior \\
        \hline
        $\cos{\theta}$ & 0.2 & 0.45 & Uniform[-1, 1]\\
        $\phi$ & 5.0 & 5.0 & Uniform[0, 2$\pi$]\\
        $\cos{\iota}$ & 0.5 & 0.5 & Uniform[-1, 1]\\
        $\log_{10}{H_0}$ & -14.4 & -14.5 & Uniform[-20, -11]\\
        $\log_{10}{M (M_{\odot})}$ & 9.5 & 9.5 & Uniform[7, 10]\\
        $\eta$ & 0.2 & 0.2 & Uniform[0.01, 0.25]\\
        $e_0$ & 0.45 & 0.5 & Uniform[0.01, 6]\\
        $\log_{10}{f_0}$ (Hz) & -8.2 & -8.2 & Uniform[-8.64, -7.52]\\
        $l_0$ & 3.5 & 4.0 & Uniform[0, 2$\pi$]\\
        $\omega_0$ & 2.0 & 2.0 & Uniform[0, 2$\pi$]\\
        $\psi$ & 2.0 & 1.5 & Uniform[0, $\pi$]\\
        \hline
        S/N & 12.3 & 5.7\\
        \hline
        \hline
    \end{tabular}
    \label{tab:injected_params}
\end{table}

\begin{figure}
    \centering
    \includegraphics[width=0.98\linewidth]{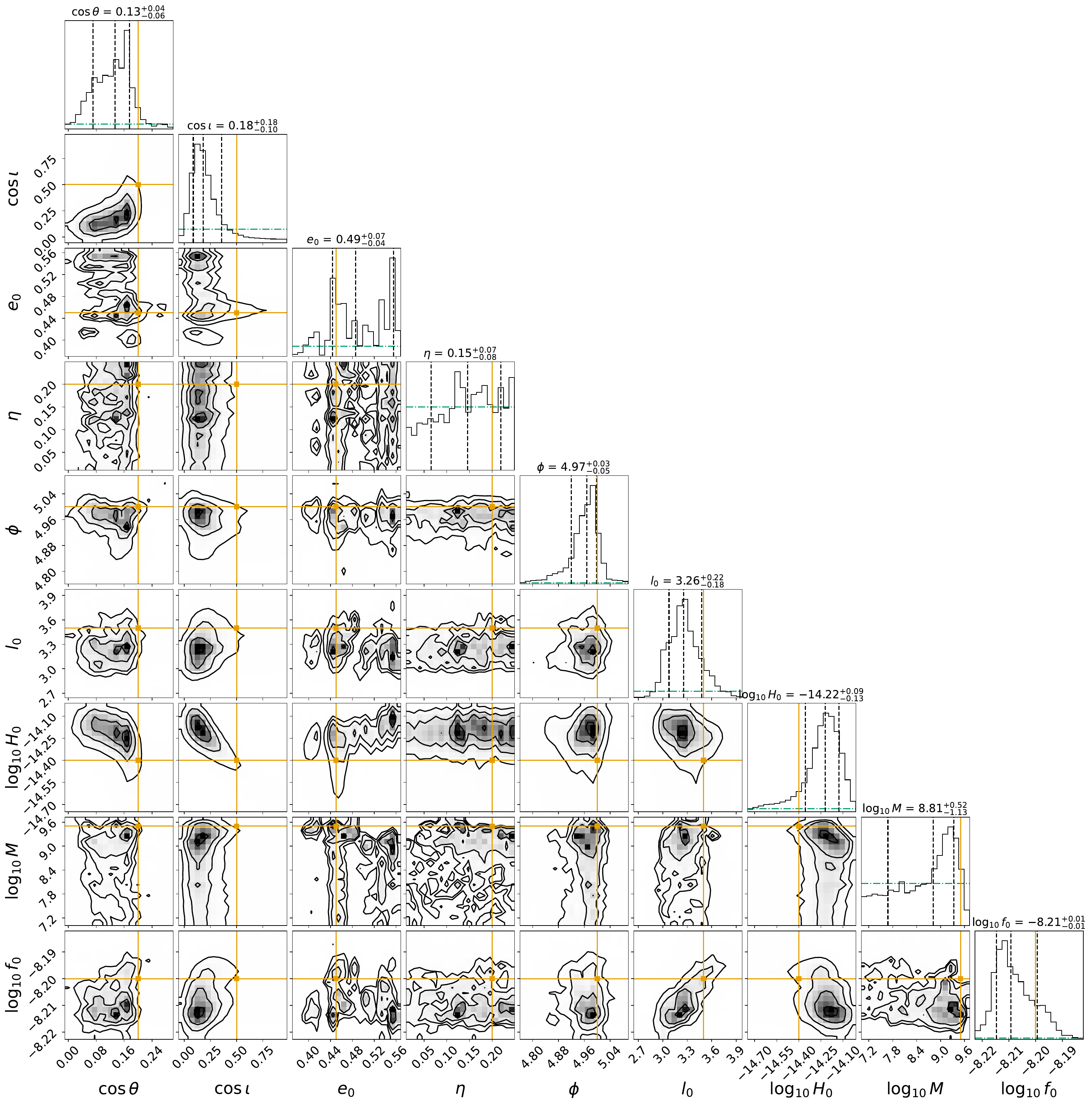}
    \caption{Corner plot showing the one- and two-dimensional marginalized posterior distributions of the SMBHB parameters for the detection analysis with an injected signal of S/N $= 12.3$. The posterior distributions are shown in black, while the green horizontal dot-dashed curves in the one-dimensional distributions indicate the corresponding prior distributions. The injected parameter values are marked in orange. The total mass ($M$) is given in units of solar masses ($M_\odot$), and the initial orbital frequency ($f_0$) is expressed in Hz.}
    \label{fig:corner_det}
\end{figure}

For the detection analysis, we inject a CGW signal with $\log_{10} H_0 = -14.4$ and the remaining binary parameters as listed in the second column of Table~\ref{tab:injected_params}, corresponding to a  S/N of approximately $12.3$. 
The injected sky location is chosen to lie in a region of the sky with good PTA sensitivity due to the relatively large number of nearby pulsars. 
We adopt a moderately high eccentricity, consistent with theoretical expectations for SMBHBs emitting GWs in the PTA frequency band. 
The reference orbital frequency is selected to lie near the most sensitive frequency of our 15-year simulated PTA dataset, while the total mass is chosen to be sufficiently large that the orbital evolution between the Earth and pulsar terms is non-negligible.
We search for this injected signal using the \texttt{QuickGWecc} pipeline and the prior distributions used for the binary parameters are shown in the last column of Table~\ref{tab:injected_params}.
Figure~\ref{fig:corner_det} shows the resulting corner plot containing the one- and two-dimensional marginalized posterior distributions of the SMBHB parameters.
The posterior distributions are shown in black, the green dot-dashed horizontal lines in the one-dimensional panels indicate the corresponding prior distributions, and the injected parameter values are marked in orange.

For all binary parameters, the recovered posterior distributions are consistent with the injected values within the 2$\sigma$ credible intervals. 
Several parameters, including the orbital frequency, sky location, and reference mean anomaly, are well constrained with narrow posterior distributions. 
This behavior is consistent with previous studies of circular SMBHBs in PTA data \citep{Petrov+2025, Charisi+2024}. 
In contrast, the posterior distributions of the signal amplitude ($\log_{10}H_0$) and cosine of the inclination angle ($\cos{\iota}$) exhibit noticeable biases, with the former tending toward larger values and the latter toward smaller values than the injected parameters. 
This behavior is primarily driven by the strong covariance between these two parameters. 
The reference eccentricity ($e_0$) and symmetric mass ratio ($\eta$) are only weakly constrained, reflecting the relatively limited information carried by the data about these parameters. 
This is consistent with the results of \citet{Agazie+2024_3C66B}, where we found similar weak constraints on $e_0$ and $\eta$ using an \texttt{enterprise}-based \citep{enterprise} targeted search pipeline for eccentric SMBHBs in PTA data.

Although the peak of the posterior distribution for the total mass ($\log_{10}M$) is consistent with the injected value, it exhibits a long tail toward lower masses.
One contributing factor to the broad posterior distribution of $\log_{10}M$ is likely the weak constraint on the pulsar-term contributions, arising from the limited precision of current pulsar distance measurements.
To investigate this further, we examined the posterior distributions of the pulsar-term parameters $d_p$, $l_p$, and $\omega_p$. 
We find that the posterior distributions of the pulsar distances are largely uninformative, consistent with previous such studies for both circular binaries using \texttt{QuickCW} pipeline \citep{Becsy+2022} and eccentric binaries using \texttt{enterprise}-based \texttt{GWecc} pipeline \citep{Agazie+2024_3C66B}.
In contrast, for a subset of pulsars, the posterior distributions of $l_p$ and $\omega_p$ while still relatively broad, are modestly informative.
We further find that these pulsars are predominantly located close to the GW source on the sky, where the GW-induced timing residuals are strongest and the pulsar-term contribution is therefore better constrained.

\begin{figure}
    \centering
    \includegraphics[width=0.5\linewidth]{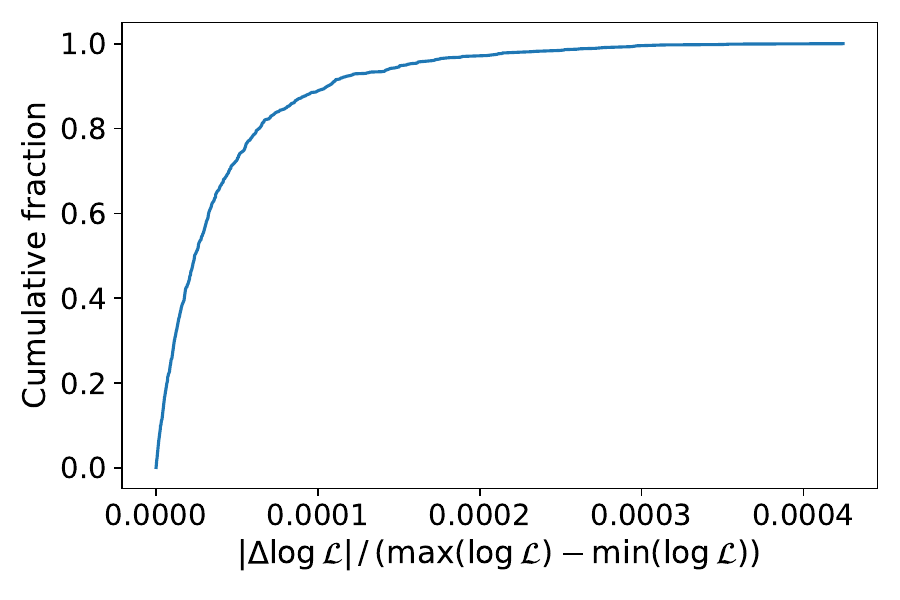}
    \caption{Cumulative distribution function of the normalized absolute difference between the log-likelihood values computed using \texttt{QuickGWecc} and \texttt{enterprise} for 1000 randomly drawn posterior samples from the detection analysis. The absolute difference is normalized by the peak-to-peak range of the log-likelihood values across the posterior samples. The small normalized differences demonstrate excellent agreement between the two likelihood implementations over the posterior region explored by the sampler.}
    \label{fig:likelihood_comparison}
\end{figure}

To validate the likelihood implementation in the \texttt{QuickGWecc} pipeline, we compare the log-likelihood values computed by \texttt{QuickGWecc} with those obtained using \texttt{enterprise} \citep{enterprise}, the standard software package for PTA data analysis, for identical sets of model parameters randomly drawn from the posterior samples of the detection analysis. 
In \texttt{enterprise}, likelihood is evaluated using the CGW-induced timing delay given by Eq.~(\ref{eq:CGW_delay}), whereas in \texttt{QuickGWecc} we use the decomposed form given in Eq.~(\ref{eq:quickGWecc_delay}).
Figure~\ref{fig:likelihood_comparison} shows the cumulative distribution of the normalized absolute difference between the log-likelihood values computed using the two implementations for 1000 randomly drawn posterior samples.
The normalization is performed using the peak-to-peak range of the log-likelihood values across the posterior samples.
We find that the normalized differences are negligible over the entire posterior region explored by the sampler, demonstrating excellent agreement between \texttt{QuickGWecc} and \texttt{enterprise} and validating the correctness of the likelihood implementation.

\subsubsection{Upper-limit analysis}

\begin{figure}
    \centering
    \includegraphics[width=0.5\linewidth]{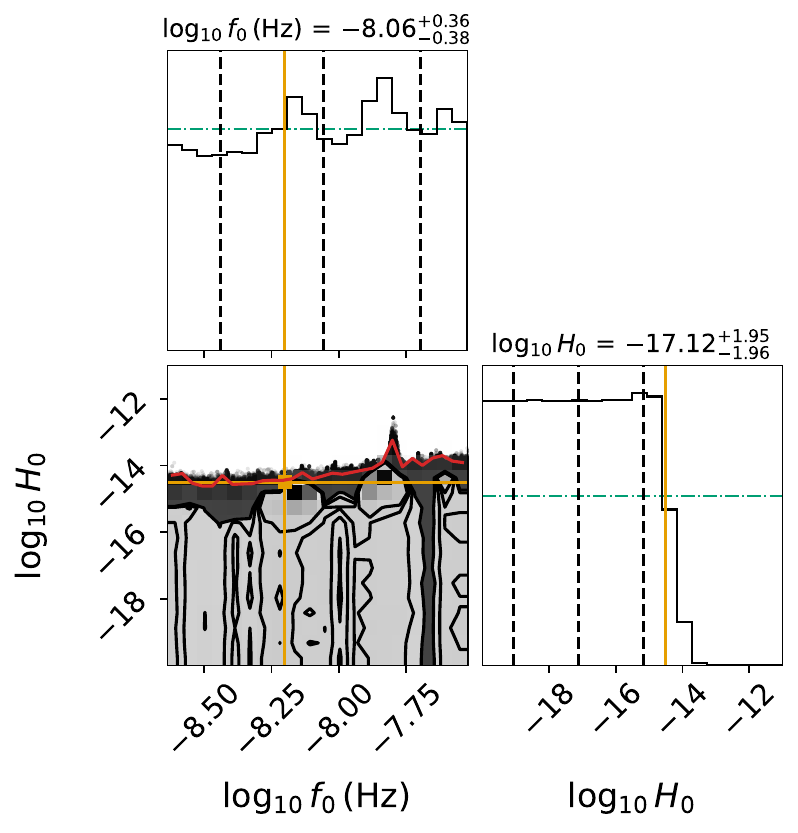}
    \caption{Corner plot showing the one- and two- dimensional marginalized posterior distributions of $\log_{10}{f_0}$ and $\log_{10}{H_0}$ for upper-limit analysis. The color scheme is same as Fig.~\ref{fig:corner_det}. The red curve in the two-dimensional distributions shows the 95\% upper limit on $\log_{10}{H_0}$ as a function of orbital frequency.}
    \label{fig:corner_ul}
\end{figure}

For the upper-limit analysis, we inject a CGW signal from an eccentric SMBHB with $\log_{10} H_0 = -14.5$, corresponding to a S/N of approximately $5.7$.
The remaining binary parameters are listed in the third column of Table~\ref{tab:injected_params}.
The resulting Savage-Dickey Bayes factor for the presence of a CGW signal from our \texttt{QuickGWecc} pipeline run is $\mathcal{B} = 0.6337 \pm 0.0007$, indicating that the signal is not detectable in this dataset.
Figure~\ref{fig:corner_ul} shows the corresponding one- and two-dimensional marginalized posterior distributions of $log_{10}{f_0}$ and $\log_{10}{H_0}$ and the posterior distributions of all other binary parameters are almost same as the prior.
Since no significant evidence for a signal is found, we compute a $95\%$ upper limit on the CGW amplitude using the prior reweighting method described in \citet{Agazie+2024_3C66B}.
The red curve in the two-dimensional distribution in Fig.~\ref{fig:corner_ul} shows the 95\% upper limit on $\log_{10}{H_0}$ as a function of the orbital frequency of the binary, while the overall upper limit is calculated to be $\log_{10}{H_0}|_{95\%}^{\rm UL} = -14.28$.

\subsection{Targeted search}

We now turn to targeted searches for CGWs from eccentric SMBHB candidates using \texttt{QuickGWecc}.
In targeted searches, prior information about the binary is available from electromagnetic observations, including the sky location $(\theta,\phi)$, luminosity distance $d_L$, and, in some cases, the orbital frequency $f_0$ \citep{NG11_2020_3C66B_targeted, Agazie+2024_3C66B, NG15_2026_114_targeted}. 
Such information can be obtained for SMBHB candidates identified through photometric periodicity, spectroscopic variability, or other electromagnetic signatures. 
To naturally incorporate these observational constraints, we use $d_L$ as an independent sampling parameter instead of $H_0$ in the targeted-search implementation, and $H_0$ is calculated using the expression $H_0 = (G M \eta x_0)/(c^2 d_L)$.
This version of the pipeline is available in the `targeted’ branch\footnote{\url{https://github.com/lanky441/QuickGWecc/tree/targeted}} of \texttt{QuickGWecc}.

\begin{figure}
    \centering
    \includegraphics[width=0.85\linewidth]{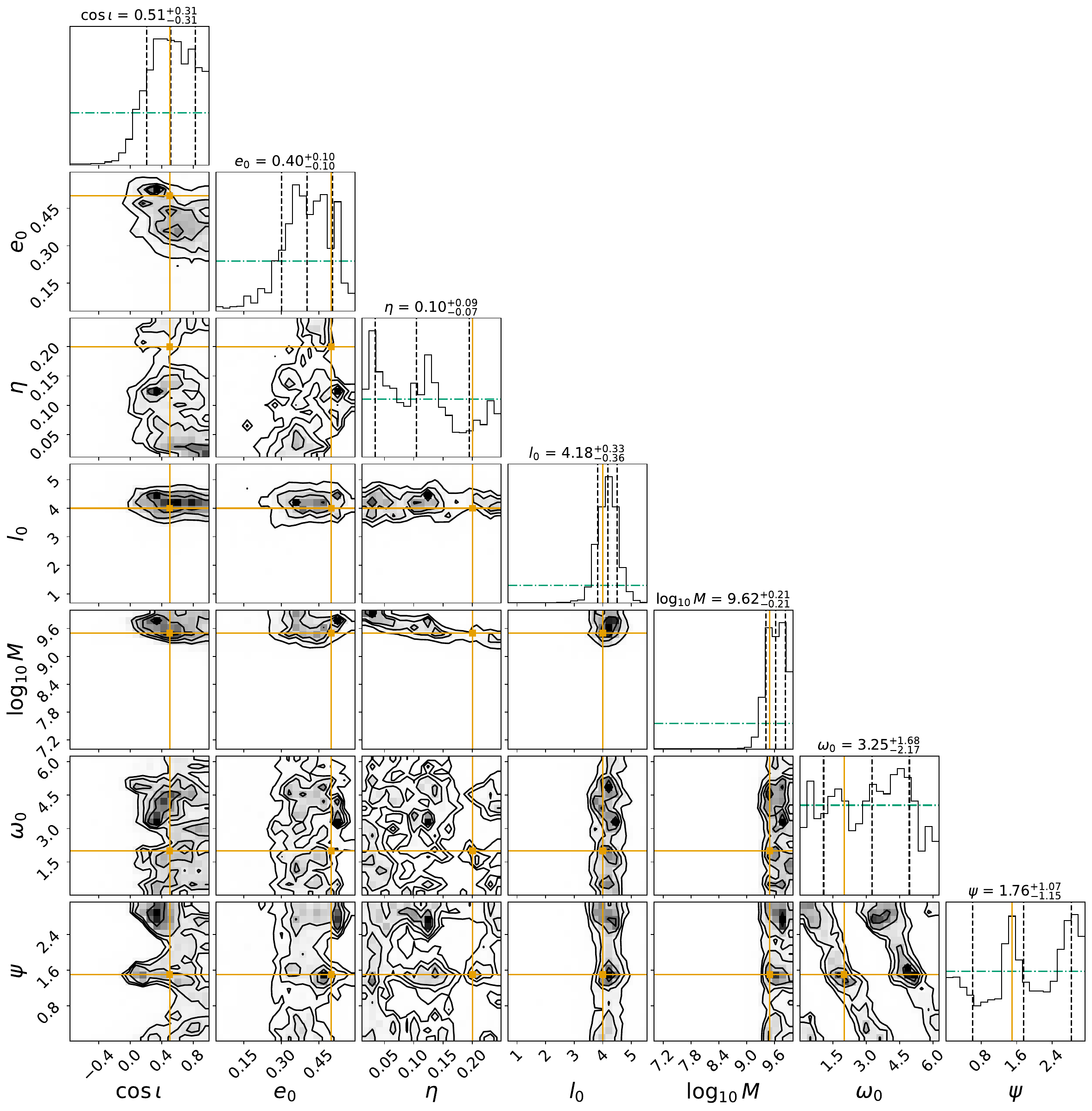}
    \caption{Same as Fig.~\ref{fig:corner_det}, but for the targeted search in the simulated dataset with S/N=5.7 used for the all-sky upper-limit analysis.}
    \label{fig:corner_targeted}
\end{figure}

To test the targeted-search implementation, we apply it to the same simulated dataset with S/N $= 5.7$ that was used for the all-sky upper-limit analysis. 
For this analysis, we fix the sky location $(\theta,\phi)$, luminosity distance ($d_L = 70.827$ Mpc; calculated from the injected values of $H_0$, $M$, $e_0$, $\eta$, and $f_0$), and orbital frequency $f_0$ to their injected values. 
The resulting Savage–Dickey Bayes factor for $\log_{10}M$ is $\mathcal{B} = 64.6 \pm 6.8$, indicating a detectable CGW signal in the targeted-search analysis.
The corresponding corner plot for the SMBHB parameters is shown in Fig.~\ref{fig:corner_targeted}.

Compared to the all-sky analysis of the same dataset, the targeted search yields a substantially higher detection significance. 
This improvement arises because fixing a subset of the binary parameters reduces the effective search volume and allows the analysis to focus on a smaller region of parameter space. 
Consequently, targeted searches provide enhanced sensitivity compared to all-sky searches and represent a promising avenue for the first multi-messenger detection of an SMBHB \citep{LiuVigeland2021}.
Such searches will be particularly relevant for future PTA datasets, including the NANOGrav 20-year and IPTA DR3 datasets, together with the growing population of electromagnetic SMBHB candidates.
At the same time, caution is warranted when interpreting candidate detections, as a significant fraction of the proposed electromagnetic SMBHB candidates may not correspond to genuine binaries. 
Robust statistical validation, including coherence tests and sky-scrambling analyses, will therefore be essential to distinguish true GW signals from statistical fluctuations or spurious electromagnetic associations \citep{Becsy+2025}.

Note that the original \texttt{QuickCW}\footnote{\url{https://github.com/nanograv/QuickCW}} implementation did not include the capability for performing targeted searches for circular SMBHBs. 
In \citet{NG15_2026_114_targeted}, \texttt{QuickCW} was used to perform targeted searches for CGWs from SMBHB candidates in the NANOGrav 15-year dataset by adopting very narrow priors on the sky location and GW frequency based on the electromagnetic observations (Agarwal et al., in prep.). 
However, the measured luminosity distance to the source was not incorporated directly into the Bayesian sampling. 
Instead, the distance information was imposed during post-processing through rejection sampling, where posterior samples were retained only if the luminosity distance inferred from the sampled values of $H_0$, chirp mass, and GW frequency was consistent with the measured distance within its uncertainty. 
Although this approach is valid, it discards a large fraction of the posterior samples, reducing the effective sample size.
By treating $d_L$ as an independent sampling parameter, the targeted implementation of \texttt{QuickGWecc} incorporates all available electromagnetic information directly into the Bayesian analysis, thereby avoiding the need for post-processing rejection sampling and making the search substantially more efficient. 
The same strategy can also be applied to targeted searches for circular SMBHBs using \texttt{QuickCW}, and has been implemented in the \texttt{targeted}\footnote{\url{https://github.com/lanky441/QuickCW/tree/targeted}} branch of \texttt{QuickCW}.

\section{Summary and Discussions}
\label{sec:summary}

In this work, we developed \texttt{QuickGWecc}, a fast Bayesian pipeline to search for CGWs from SMBHBs in relativistic eccentric orbits in PTA datasets. 
The pipeline extends the \texttt{QuickCW} framework \cite{Becsy+2022}, originally developed for circular binaries, to eccentric SMBHBs motivated by theoretical studies suggesting that binaries emitting GWs in the PTA frequency band may retain significant eccentricity.
The computational efficiency of \texttt{QuickGWecc} is achieved by separating the model parameters into shape and projection parameters, allowing repeated updates of the computationally inexpensive projection parameters while keeping the shape parameters fixed.

For a NANOGrav 15-year-like dataset, the average likelihood evaluation time for a single update to the projection parameters in \texttt{QuickGWecc}, measured on a MacBook Pro equipped with an Apple M4 chip, is approximately $41~\mu\mathrm{s}$, while a shape parameter update requires approximately $580~\mathrm{ms}$. 
For comparison, the average likelihood evaluation time in \texttt{enterprise} is approximately $559~\mathrm{ms}$. 
Although the computational cost of updating the shape parameters in \texttt{QuickGWecc} is comparable to that of an update in \texttt{enterprise}, the extremely fast projection-parameter updates allow thousands of proposals to be explored in negligible amount of time.
Combined with a Metropolis-within-Gibbs sampler and a multiple-try MCMC scheme, this enables efficient exploration of the high-dimensional parameter space relevant for eccentric SMBHB searches.


We validated the likelihood implementation by comparing the likelihood values computed using \texttt{QuickGWecc} with those obtained from \texttt{enterprise}, finding excellent agreement between the two methods. 
Using simulated PTA datasets containing white noise, intrinsic pulsar red noise, a stochastic GW background, and CGWs from an eccentric SMBHB, we demonstrated the performance of the pipeline for both detection and upper-limit analyses. 
We also introduced a targeted-search implementation optimized for searching CGWs from electromagnetic SMBHB candidates.

Although the current pipeline demonstrates the feasibility of efficient eccentric CGW searches in PTA data, we plan to address several remaining limitations in future works. 
At present, the pipeline assumes that DM variations are modeled using the DMX formalism incorporated into the timing model. 
However, future analyses of real PTA datasets would benefit from more flexible noise modeling approaches, such as replacing the \texttt{DMX} model with GP models for DM variations. 
Similarly, incorporating GP models for solar-wind-induced delays, particularly for pulsars located close to the ecliptic plane, would improve the characterization of low-frequency chromatic noise contributions \citep{Larsen+2026_NG15_CNM}. 
This will become increasingly important with the inclusion of lower radio frequency observations from telescopes like uGMRT, CHIME, and LOFAR in future PTA datasets.

Another important limitation of the current implementation is the treatment of the stochastic GWB as a CURN. 
The present version of \texttt{QuickGWecc} does not directly support searches in the presence of a HD correlated GWB. 
However, likelihood reweighting techniques can be used to approximately recover posteriors corresponding to an HD-correlated background model \citep{Hourihane+2023}. 
Extending the likelihood framework to directly incorporate HD correlations is an important direction for future development and will be particularly relevant for analyses of next-generation PTA datasets.

The upcoming NANOGrav 20-year dataset and the third International Pulsar Timing Array data release (IPTA DR3) will provide significantly improved sensitivity to nanohertz GWs through longer baselines, larger pulsar samples, and improved timing precision. 
These datasets are expected to enhance the prospects for detecting both the stochastic GW background at the $5\sigma$ level and the loudest individual CGW sources. 
The computational efficiency of \texttt{QuickGWecc} makes it well suited for all-sky searches for eccentric SMBHBs in these larger datasets, where conventional eccentric searches may become prohibitively expensive.
Furthermore, the targeted-search implementation provides a framework for systematic searches of electromagnetic SMBHB candidates identified through periodic variability, spectroscopic signatures, or multi-wavelength observations.

The recent development of the \texttt{discovery} pipeline has further advanced rapid parameter estimation for PTA datasets \citep{Vallisneri+2025_discovery}.
Instead of relying on traditional MCMC methods, \texttt{discovery} employs stochastic gradient-based Bayesian variational inference and minimizes the Kullback–Leibler divergence to efficiently approximate the posterior distribution. 
Extending \texttt{discovery} to incorporate the CGW signal model for SMBHBs in relativistic eccentric orbits could further reduce the computational cost of eccentric CGW searches, making analyses of increasingly sensitive PTA datasets even more efficient.
The detection of CGWs from an eccentric SMBHB would have important implications for understanding SMBHB formation and environmental interactions during binary evolution.
Measurements of eccentricity in the PTA band can constrain dynamical processes such as stellar scattering, gas interactions, and triple-body encounters that govern the orbital evolution of SMBHBs.
Therefore, \texttt{QuickGWecc} provides not only a computational tool for future PTA analyses but also a framework for probing the astrophysics of SMBHB evolution and enabling multi-messenger studies in the nanohertz GW regime.

\bibliography{quickgwecc}

\end{document}